\documentclass[fleqn,usenatbib]{mnras}

\usepackage{newtxtext,newtxmath}

\usepackage[T1]{fontenc}

\usepackage{soul}
\DeclareRobustCommand{\VAN}[3]{#2}
\let\VANthebibliography\thebibliography
\def\thebibliography{\DeclareRobustCommand{\VAN}[3]{##3}\VANthebibliography}

\usepackage{graphicx}	
\usepackage{amsmath}	

\title[Ion abundances in the plasma tail of 3I/ATLAS]{Ion abundances in the plasma tail of 3I/ATLAS show that it is N$_2$-rich}

\author[L. Ferellec]{
L. Ferellec,$^{1}$\thanks{E-mail: lea.ferellec@northumbria.ac.uk}
C. Opitom,$^{2}$
C. Snodgrass$^{2}$
\\
$^{1}$Northumbria University, Ellison Pl, Newcastle upon Tyne NE1 8ST, United Kingdom\\
$^{2}$Institute for Astronomy, University of Edinburgh, Royal Observatory, Blackford Hill, Edinburgh EH9 3HJ, United Kingdom\\
}

\date{Accepted XXX. Received YYY; in original form ZZZ}

\pubyear{\the\year{}}

\begin{document}
\label{firstpage}
\pagerange{\pageref{firstpage}--\pageref{lastpage}}
\maketitle

\begin{abstract}
Interstellar objects allow us to compare the early planetary formation process in the Solar System and around other stars. The second and third interstellar objects, 2I/Borisov and 3I/ATLAS, showed visible signs of activity, making it possible to assess their composition via spectroscopic analysis of their gas coma. While neutral volatiles surrounding comets are easily probed by common spectrographs, ion emissions emanate mostly from the tail, making them easier to detect with integral-field units. We present WHT/WEAVE-LIFU observations of 3I/ATLAS, performed post-perihelion on 2025 November 30 and 2025 December 02. We simultaneously detect the ions N$_2$$^+$, CO$^+$, CO$_2$$^+$, H$_2$O$^+$ and CH$^+$ located in an anti-solar tail. Using common fluorescence factors, we determine abundance ratios of these species in the tail. In particular, the N$_2$$^+$/CO$^+$ ratio allows us to constrain the lower-limit N$_2$/CO$>$0.023$\pm$0.001, indicating that 3I/ATLAS is N$_2$-rich compared to Solar System comets, a property which can be linked to cold formation conditions. Abundance ratios between other ions cannot directly be used to constrain abundances of neutrals but seem consistent with the fact that 3I/ATLAS is enriched in hypervolatiles. We also investigate whether these ratios vary along the tail, and only detect a marginal decrease of CH$^+$ with cometocentric distance.
\end{abstract}




\section{Introduction} 
Comets are pristine remnants of the early stages of planetary formation. Their ices, unaltered, bear witness to the physical and chemical conditions of the environment in which they formed. Local comets are therefore key to constrain the formation and evolution of the Solar System. However, in recent years, small bodies visiting from different planetary systems have been discovered and have shown comet-like activity. These  objects are an opportunity to compare planetary formation in the Solar System and in other stellar environments, and to assess the diversity in our Galaxy.

The first interstellar object 1I/'Oumuamua was discovered on 2017 October 19, already on its way out of the Solar System \citep{2017Natur.552..378M}. No visual signs of activity (gas or dust) were detected, although its trajectory showed signs of non-gravitational acceleration, consistent with outgassing \citep{2018Natur.559..223M}.
In contrast, the second interstellar object 2I/Borisov, discovered on 2019 August 29\footnote{Minor Planet Electronic Circulars \href{https://www.minorplanetcenter.net/mpec/K19/K19RA6.html}{ 2019-R106}}, displayed outgassing with a composition consistent with Solar System comets \citep[e.g.][]{2021MNRAS.502.3491A,2021A&A...650L..19O,2026PSJ.....7...88D}  although enriched in hypervolatile CO \citep{2020NatAs...4..867B, 2020NatAs...4..861C}. 
The third interstellar object was highly anticipated until its discovery on 2025 July 1. 3I/ATLAS (hereafter 3I) was identified as cometary on 2025 July 2\footnote{Minor Planet Electronic Circulars \href{https://minorplanetcenter.org/mpec/K25/K25N12.html}{2025-N12}}, inbound for a perihelion passage at r$_h$=1.36au on 2025 October 29.

Studies from ground-based and space-based facilities extensively constrained the composition of the coma of 3I during its apparition. 
Abundances of the main volatiles H$_2$O, CO$_2$ and CO (infrared observations) showed that 3I was enriched in CO$_2$ and CO pre perihelion \citep{2025ApJ...991L..43C,2026ApJ..1000L..52L}, but that CO$_2$ decreased post perihelion \citep{2026arXiv260306911C,2026RNAAS..10...26L}. 
From optical observations, the composition also evolved from carbon-depleted pre-perihelion to typical post perihelion \citep{2025ApJ...995L..34R,2026A&A...706A..43H,2026arXiv260116983H}. 
3I was found to have high abundances of Fe and Ni, with a Ni/Fe ratio larger than what is usually measured for solar system comets, that decreased as the comet approached the Sun pre-perihelion \citep{2026A&A...706A..43H,2025ApJ...995L..34R}. Post-perihelion measurements of the Ni/Fe ratio match measurements made pre-perihelion at a similar heliocentric distance \citep{2026arXiv260116983H}. 3I is surprisingly poor in NH$_3$, as indicated by the lack of detection of NH$_2$ at optical wavelengths \citep{2026ApJ..1000L..60K}.

Measurements of C, N, and H isotopic ratios have provided clues on the formation environment of 3I. \citet{2026arXiv260307026S} and \citet{2026arXiv260306911C} measured a strong deuterium enrichment, more than one order of magnitude higher than what is measured for Solar System comets. This indicate that the object probably formed in a cold environment, with temperatures  $\leq$30 K. This is reinforced by a measurement of $^{14}$N/$^{15}$N that points toward a formation in the outskirts of the protoplanetary disk \citep{2026arXiv260307187O}. \citet{2026arXiv260307187O} and \citet{2026arXiv260306911C} also report a $^{12}$C/$^{13}$C higher than what is measured for solar system comets, consistent with a formation in a low metallicity environment early on in the Milky Way history. 

{\color{black}A field still understudied for interstellar objects is their ion production. Cometary ions are produced from neutral species via various processes such as photo-ionization, electron-impact or charge exchange. They interact with the incoming solar-wind and interplanetary magnetic field, creating an ion tail that can span multiple astronomical units, with dynamics distinct from the dust tail. H$_2$O$^+$ and CO$^+$ are the most commonly seen cometary ions, abundant and creating bright emissions in the visible range, allowing ion tails to often be visible in broadband imaging. Despite decades of remote and in-situ investigations of comet plasma, there is still a disconnect between the sophisticated photo-chemical and magneto-hydrodynamical models of comet plasma and their observations \citep{2024come.book..501B,2024come.book..543G}. So far for interstellar objects, the plasma tail of 2I/Borisov was detected via scintillations during a radio occultation \citep{2022PSJ.....3..266M} but individual ion species were not detected spectrally.

Spectroscopic detections of ions can provide information on their neutral parent species which cannot be detected in the same wavelength range but hold valuable insight on the history of the comet: For instance in the case of 3I, N$_2$ probed via the proxy of N$_2$$^+$ can help complete the picture of its seemingly-low nitrogen content. The abundance of N$_2$ in comparison to CO is also correlated to the formation temperature of a comet} \citep{2007Icar..190..655B}. 
As ion emissions can easily be missed by slit-spectrographs if they are not aligned with the plasma tail, integral-field units (IFUs) are a powerful resource, able to disentangle spatially the different structures from which emissions originate. Indeed, CO$^+$ and H$_2$O$^+$ ions were detected in 3I with VLT/MUSE as early as 2025 September (Ferellec et al. in preparation). In this letter, we present post-perihelion observations of 3I using the WEAVE Large-IFU (LIFU) on the William Herschel Telescope (WHT, Spain), allowing us to detect the ions N$_2$$^+$, CO$^+$, CO$_2$$^+$, H$_2$O$^+$ and CH$^+$. 

\begin{figure*}
    \centering
    \includegraphics[width=1\linewidth]{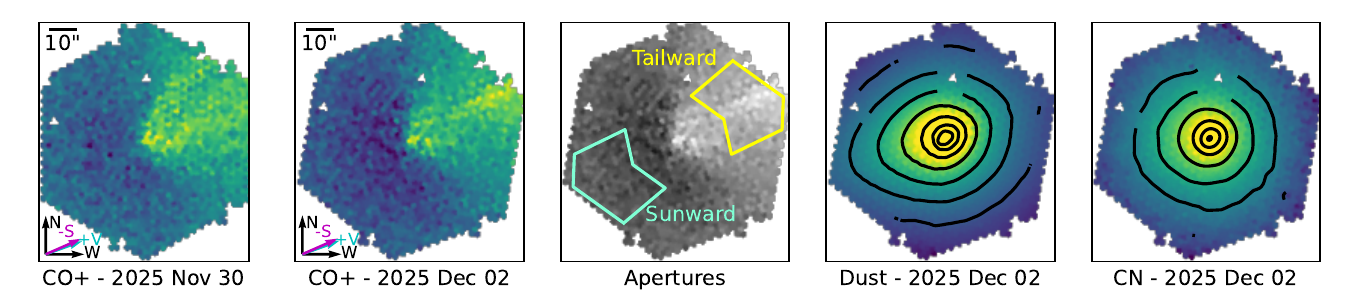}
    \caption{CO$^+$ ion maps from 2025 November 30 (first panel) and 2025 December 02 (second panel). {\color{black}At $\Delta$=1.9au, the 10" scalebar corresponds to $\sim$13800km projected at the comet.} The ion tail is visible, close to the expected anti-solar direction. Arrows indicate the anti-solar direction ("-S") and the direction of motion ("+V"). Apertures to extract tailward and sunward spectra are illustrated in the third panel. A dust map (fourth panel) and a CN map (fifth panel) from December 02 show that the dust coma extends into a sunward tail while the gas coma looks more symmetrical. Qualitative contour plots are shown, based on the images smoothed through a Gaussian filter and logarithmic contour levels.}
    \label{fig:aperture}
\end{figure*}

\begin{figure*}
    \centering
    \includegraphics[width=1\linewidth]{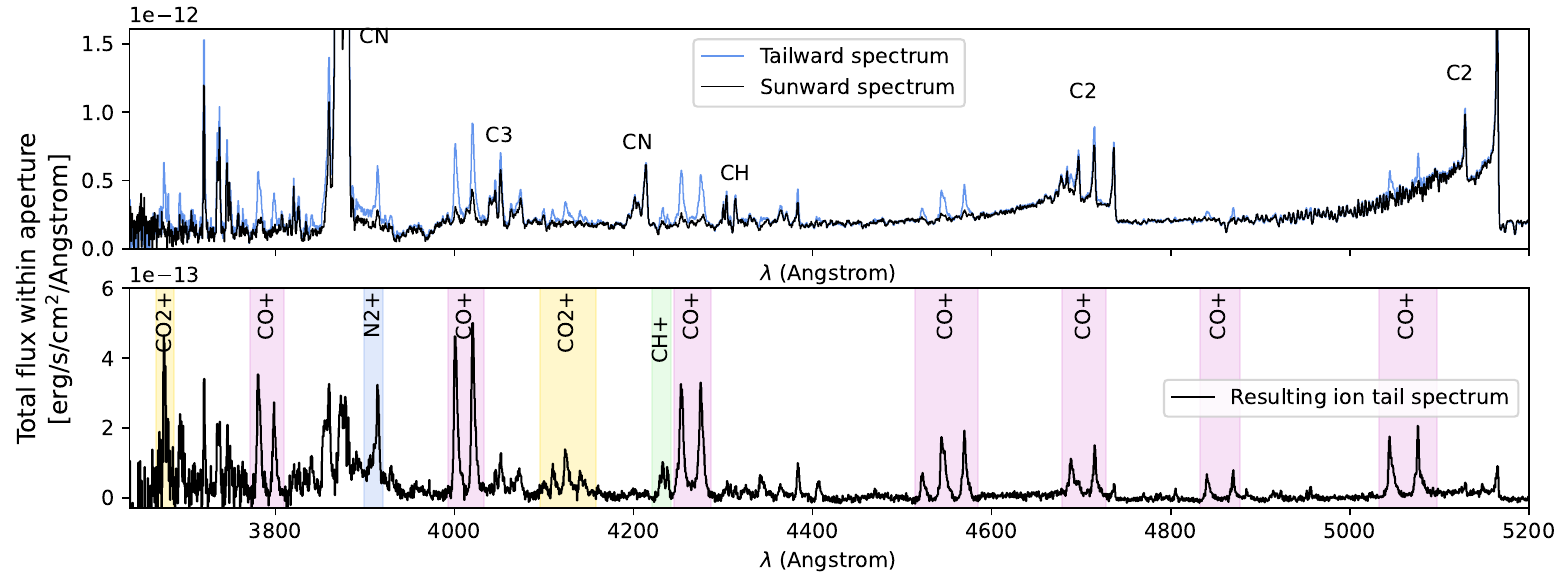}
    \caption{\textbf{Top:} Example of tail-side and antitail-side spectra from the LIFU blue arm, the locations of the apertures having been optimised to have similar gas and dust spectral components. Emission regions of the main neutral volatiles are labelled on the spectrum. Some emission lines are visible that are stronger in the tailside spectrum, showing that they are due to ions in the plasma tail. \textbf{Bottom:} Isolated ion emissions resulting from the subtraction of the antitail-side spectrum from the tail-side spectrum. The most prominent emission lines are labelled.}
    \label{fig:ions}
\end{figure*}
\section{Data and methods}

\subsection{Data}
3I was observed with WEAVE LIFU \citep{2024MNRAS.530.2688J} on 2025 November 30 and 2025 December 02, when the comet was at $r_{h}=1.8$au and $\Delta=1.9$au.
LIFU offers a large FOV of 90" by 78", sampled by a collection of individual 2.6" diameter fibres. Each epoch represents a one-hour observing block made up of 3 consecutive 1020-seconds exposures, {\color{black}with differential tracking to follow the target}. The low resolution "blue+red" mode was used, providing a resolving power R$\sim$2500 over a full range of 3660-9590\r{A}\, (3660-5900\r{A}\, for the blue arm, and 5860-9590\r{A}\, for the red arm). The WEAVE data reduction pipeline outputs {\color{black}L1} data cubes for each of the blue and red arms of the LIFU, which we calibrate in flux using the provided sensitivity functions {\color{black}\citep[WEAVE data processing and files are described in][or \href{https://camcead.ast.cam.ac.uk/weave/data_model/docs/WEAVE-ICD-027\%20WEAVE\%20Science\%20Processing\%20and\%20Analysis\%20ICD.pdf}{WEAVE Science Processing and Analysis manual}]{2024MNRAS.530.2688J}}.

Figure \ref{fig:aperture} shows intensity maps of the CO$^+$(1-1) emission line in the field of view for each epoch. These are obtained by, for each pixel, summing up the flux within the emission range and subtracting a linear interpolation of the surrounding continuum (which for this line is due to dust and C$_2$ emissions). {\color{black}This line was chosen to produce our qualitative CO$^+$ maps because, although it overlaps with a strong C$_2$ band, the shape of this band is more easily approximated by a linear continuum in comparison to the regions neighbouring other bright CO$^+$ lines.} Figure \ref{fig:aperture} also shows a map of the CN(1-0) emission line, obtained similarly with the dust continuum removed, and a map of the dust continuum emissions {\color{black}(660-680nm, a region free of emission lines)}.  
At both epochs the ion tail is clearly visible {\color{black} in the CO$^+$ maps}. The tail orientation matches roughly with the projected anti-solar direction (which would be 67$^\circ$ degrees westward from the North), although there is some variability its morphology, looking more diffuse on November 30, maybe indicative of changes in the solar-wind intensity or direction. 
The map of dust emissions show that dust coma seems slightly extended sunward, in agreement with other reports of a sunward tail seen post-perihelion \citep[e.g.][]{2026MNRAS.550g1164M} due to the anti-sunward trajectory of 3I. {\color{black}In comparison the map of CN emissions appears more symmetrical, although} \citet{2026arXiv260116983H} {\color{black}reported some asymmetries in the inner coma, observing a subsolar jet. }

\subsection{Ion spectra}

As shown on Figure \ref{fig:aperture}, we define a polygonal aperture sampling the ion tail. This "elbow" shape of the aperture aims at maximizing the tail coverage while avoiding the inner coma with stronger dust and gas signals. At the time of observation, a perfectly anti-solar ion tail would have been separated from the line of sight by $\sim 30^{\circ}$, meaning that the aperture covers physical nucleocentric distances ranging between $\sim5\times10^{4}$km and $\sim1.1\times10^{5}$km projected along this tail {\color{black}axis}. The tail-side spectrum extracted in this aperture is a superposition of the dust continuum (reflected solar light), emission lines from the neutral gas and emission lines from the ions. In order to disentangle the ion emissions from the other components, we consider a "mirrored" aperture on the sunward side of the comet, which should probe similar amounts of gas and dust as the tailward aperture but not contain the ion tail. {\color{black}We manually refine an optimal symmetry centre which allows to best match the tailward and sunward spectra in order to subtract the gas and dust from the tail side. As the dust and gas show different spatial distributions, We prioritise matching the dust continuum, resulting in some residual gas emissions, but these will either fall outside of our selected analysis ranges (see section \ref{sec:lines}) or will be easily approximated by a linear local continuum.}

\begin{figure*}
    \centering
    \includegraphics[width=0.45\linewidth]{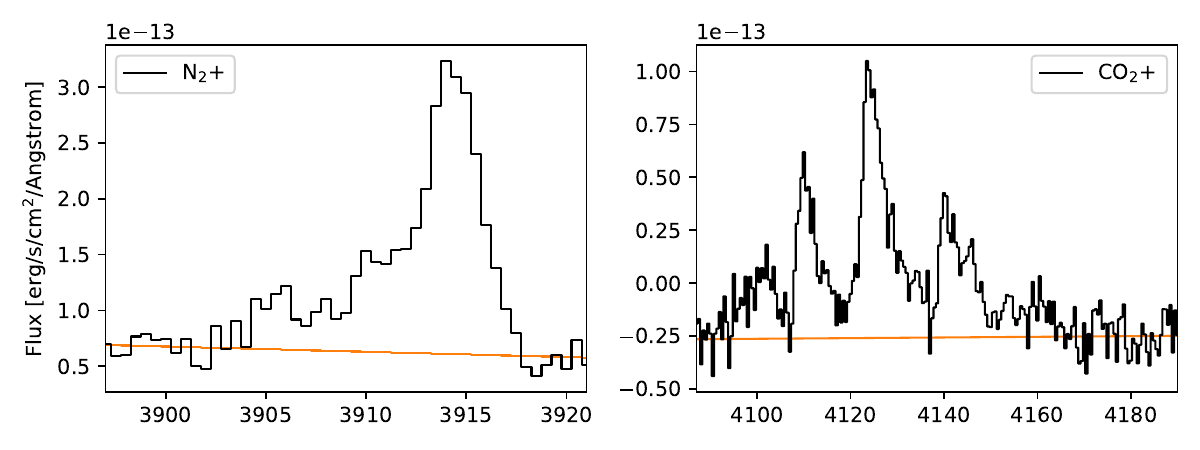}
    \includegraphics[width=0.8\linewidth]{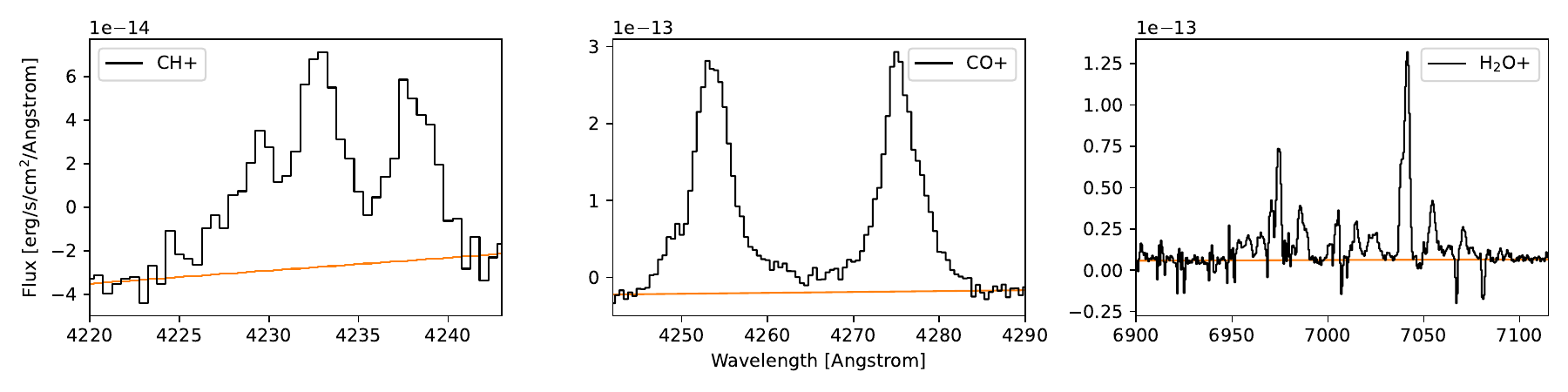}
    \caption{Emission bands of N$_2$$^+$, CO$_2$$^+$. CH$^+$, CO$^+$ and H$_2$O$^+$ used to determine abundance ratios. The ion spectrum is shown before local continuum subtraction. Orange lines show the local continuum models.}
    \label{fig:lines}
\end{figure*}

Figure \ref{fig:ions} (top panel) shows the blue arm spectra extracted in both tailward and sunward apertures. For conciseness we omit showing the red arm spectrum, but it only features emission lines of CN and of H$_{2}$O$^+$. {\color{black}One of the  H$_{2}$O$^+$ line system is shown on Figure \ref{fig:lines}.} Strong emission lines from the main neutrals are present: CN, C$_{3}$, CH, C$_{2}$. As previous spectroscopic studies reported, in the red-arm (not shown) spectrum we do not detect any NH$_{2}$ emissions. 

The ion tail spectrum (Fig. \ref{fig:ions} bottom panel), resulting from the subtraction of the anti-tail spectrum from the tail-side spectrum, shows numerous emission lines, which we identified as CO$_2$$^+$, N$_2$$^+$, CO$^+$, CH$^+$ and H$_2$O$^+$. One should note that {\color{black}the brightest} ion signatures are faintly visible in the anti-tail spectrum too, indicating that the ions are produced in the entire coma. An equivalent contribution from this extended ion source should also be present in the tail-side aperture\footnote{{\color{black}Asymmetrical neutral production was detected in the inner coma of 3I \citep[e.g.][]{2026arXiv260320460R} but such asymmetries are attenuated farther in the coma and assumed to be negligible here.}}, leaving only the ion flux from the tail in the subtracted spectrum. Some residuals from neutral volatiles are visible, e.g. the C$_3$ bands around 405nm, the head of C$_2$ band at 516nm, or around the CN band at 386nm, but as previously mentioned these should not impact the ion emissions of interest (section \ref{sec:lines}).

\subsection{Measuring abundance ratios}
The integrated flux $F_{X,\lambda_X}$ in a given emission line $\lambda_X$  depends on the column density $N_X$ of molecules X in the aperture and on the fluorescence rate of one molecule for this given transition $g_{X,\lambda_X}$. Abundance ratios in the portion of the tail probed by the aperture can be determined as follows:
\begin{equation}
    \frac{N_X}{N_Y}=\frac{F_{X,\lambda_X}}{F_{Y,\lambda_Y}}\frac{g_{Y,\lambda_Y}}{g_{X,\lambda_X}}
\end{equation}

Using the isolated ion spectra, we approximate the remaining continuum residuals around each emission line by a polynomial function. We then calculate the integrated flux in the continuum-removed emission band. We assess the error on the integrated flux as $RMS \times \Delta\lambda / \sqrt{N_{pix}}$ where $RMS$ is the standard deviation of the noise in the regions surrounding the line, $\Delta\lambda$ is the integration range, and $N_{pix}$ is the number of data points in the integration range. To include error induced by the continuum removal step, we add to this uncertainty 5\% of the integrated flux of the continuum model. These uncertainties on fluxes are propagated to determine an uncertainty $\delta$ on each measured ratio.

To account for the variability linked to our aperture selection, we repeat these calculations as we shift the aperture by a few pixels in multiple directions. Our final ratio measurement for each epoch is the average of the ratios for all aperture shifts.
Our final uncertainty is the geometric sum of the standard deviation of ratios for all aperture shifts and the average $\delta$ value. It is important to note that we do not consider uncertainties on the selected fluorescence factors. For example, \citet{1993ApJ...407..402L} consider a 50\% uncertainty on their calculated CH$^+$ fluorescence factor, but as our g-factors for different ions originate from different sources we cannot consistently assess and include such uncertainties. Therefore, when comparing to other measurements, one should keep in mind that the chosen g-factor proportionally affects the calculated abundance ratios.

\subsection{Emission lines of interest}
\label{sec:lines}

We use the emission bands and fluorescence factors listed in Table \ref{tab:gfactor}. The emission lines are illustrated in Fig. \ref{fig:lines}, showing that they are well detected.
The \textbf{N$_2$$^+$}(0-0) band at is located at 3915\r{A}. The shape of the band, including the small peak at 3905\r{A}, seems consistent with the fluorescence model of \citet{2022A&A...661A.131R} superimposed onto a linear residual continuum. 
For \textbf{CO$_2$$^+$} we use the line complex located between 4087 and 4212\r{A}. 
The other CO$_2$$^+$ band visible at 3675\r{A} suffers from the lower detector sensitivity in this region. Our attempt at deriving abundances from the 3675\r{A}$ $ line using the g-factor from \citet{1999EP&S...51..139K} gave similar results for November 30 but resulted in less consistency between the two epochs, therefore we chose not to use it.
{\color{black} For \textbf{CO$^+$} we use the two (2-0) lines at 4255\r{A} and 4275\r{A}, as they are bright and located in a region free of neutral emissions in between the CN(0-1) and CH(0-0) bands. Using CO$^+$(2-0) also allows consistent comparison with the abundance ratios recalculated by \citet{2023MNRAS.524.5182A}}.
We also use the \textbf{CH$^+$}(0-0) band located next to it, around 4230\r{A}. We expect minimal overlap between the CH$^+$ and CO$^+$ features.
Finally for \textbf{H$_2$O$^+$} we use the (6-0) complex near 7000\r{A} \footnote{{\color{black}Some narrow negative peaks are visible amongst the H$_2$O$^+$(6-0) band system (Fig. \ref{fig:lines}). These outlier values seem to be instrumental artefacts impacting this spectral region. Their intensities, sometimes positive, vary differently between fibres and wavelengths. 
As our uncertainty estimation already accounts for spectral noise and spatial fluctuations, we consider these artefacts negligible for our broad interpretations of H$_2$O$^+$ abundances. In contrast \cite{1987ApJ...315L.147L} estimates a 25$\%$ uncertainty on their water ion g-factors, and we will compare our measurements to H$_2$O measurements that might also suffer from biases (extended sources).}}, as the (8-0) complex at 6200\r{A} might suffer from contamination by CO$^+$. We attempted to use the (8-0) band using the g-factor given by \citet{1987ApJ...315L.147L} but it resulted in column densities $\sim30\%$ larger.

\begin{table}
    \centering
    \begin{tabular}{|c|c|c|c|} \hline
        Emission & Range & $g$-factor at 1au & Source for  \\ 
        line & [\r{A}] & [photon/molecule/s] &  the $g$-factor \\ \hline
        N$_2$$^+$ (0-0) & 3897-3921 & $4.9\times10^{-2}$ (at 3915\r{A}) & 1 \\
        CO$_2$$^+$ $^{\star}$ & 4087-4212 & $1.4\times10^{-3}$ (at 4125\r{A}) & 2 \\
        CH$^+$ (0-0) & 4220-4243 & $2.8\times10^{-2}$ (at 4235\r{A}) & 3 \\
        CO$^+$ (2-0) & 4242-4290 & $3.55\times10^{-3}$ (at 4265\r{A}) & 4\\
        H$_2$O$^+$ (6-0) & 6900-7115 & $6.5\times10^{-3}$ (at 7040\r{A}) & 5\\ \hline
    \end{tabular}
    \caption{Emission bands and fluorescence factors used to derive abundance ratios. For each g-factor we specify the wavelength that we assumed to convert emitted photons to energy. $^\star$For CO$_2$$^+$, we calculated an equivalent g-factor based on all the transitions that fall within the wavelength range, wich are v''-v'=1 for bands 3 and 4 according to the labelling system of \citet{1999EP&S...51..139K}. Sources for g-factors are 1-\citet{2022A&A...661A.131R} ; 2-\citet{1999EP&S...51..139K} ; 3-\citet{1993ApJ...407..402L} ; 4-\citet{1986ApJ...302..477M} ; 5-\citet{1987ApJ...315L.147L}.}
    \label{tab:gfactor}
\end{table}

\section{Results and discussion}
\subsection{Bulk abundance ratios}

Table \ref{tab:results} presents the abundance ratios derived from the observations. {\color{black} averaged for both epochs. Ratios derived at each epoch are given in appendix \ref{sec:appendix} and agree within the uncertainties.}

\begin{table}
\begin{tabular}{|c|c|c|c|c|c|}
\hline
\multicolumn{6}{|c|}{\textbf{Abundance ratios A/B [\%] (average of both epochs)}} \\ \hline
 & A=N$_2$$^+$ & A=CO$^+$ & A=CO$_2$$^+$ & A=H$_2$O$^+$ & A=CH$^+$ \\ \hline
B=N$_2$$^+$  &  100 & 4300$\pm$200 & 6000$\pm$600 & 1500$\pm$100 & 110$\pm$10  \\ \hline
B=CO$^+$  &  2.3$\pm$0.1 & 100 & 140$\pm$10 & 35$\pm$2 & 2.5$\pm$0.2  \\ \hline
B=CO$_2$$^+$  &  1.7$\pm$0.2 & 72$\pm$6 & 100 & 26$\pm$3 & 1.8$\pm$0.9  \\ \hline
B=H$_2$O$^+$  &  6.6$\pm$0.4 & 290$\pm$10 & 380$\pm$40 & 100 & 7.2$\pm$0.7  \\ \hline
B=CH$^+$  &  90$\pm$7 & 3900$\pm$300 & 5400$\pm$600 & 1400$\pm$100 & 100  \\ \hline
\end{tabular}
\caption{Measured abundance ratios of species A/B given in percents.}

\label{tab:results}
\end{table}

\subsubsection{N$_2$/CO} 
\begin{figure}
    \centering
    \includegraphics[trim={0 0.8cm 1.5cm 0.6cm},clip,width=1\linewidth]{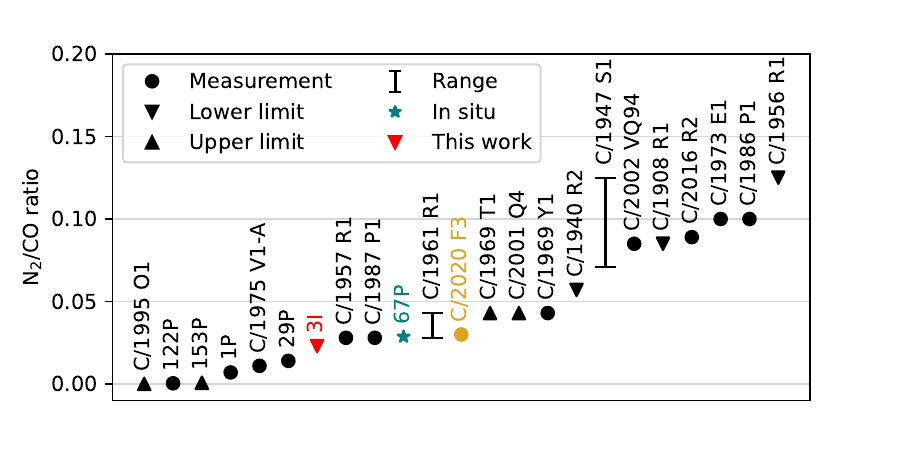}
    \caption{N$_2$/CO ratios measured in 3I (this work, in red), {\color{black}in C/2020 F3 \citep[][in gold]{2025A&A...701A.161A}}, in other N$_2$-rich comets \citep[as summarized by ][in black]{2023MNRAS.524.5182A}, and in-situ measurement for 67P from \citep[][in teal]{2019MNRAS.489..594R}. 
    {\color{black} The depletion compared to a protosolar ratio N$_2$/CO$\sim$0.15 can be used to estimate the formation temperature of Solar System comets.}
    }
    \label{fig:N$_2$coratios}
\end{figure}

N$_2$ and CO having similar branching ratios for photoionization into N$_2$$^+$ and CO$^+$, N$_2$$^+$/CO$^+$ can usually be used as a proxy for N$_2$/CO. 
In our case, given the high abundance of CO$_2$ in the coma which could indirectly produce CO$^+$, N$_2$$^+$/CO$^+$ can be treated as a lower-limit for N$_2$/CO trapped in the ice. In this context, we find that the N$_2$/CO ratio exhibited by 3I is N$_2$/CO$>$0.023±0.001. This value is similar to the few N$_2$-rich comets of the Solar System for which N$_2$/CO ratios have been measured, listed by \citet{2023MNRAS.524.5182A} and illustrated by Figure \ref{fig:N$_2$coratios}. {\color{black} Among these N$_2$-rich comets, C/2016 R2 was also depleted in NH$_3$ \citep{2019AJ....158..128M} and  NH$_2$ \citep{2019A&A...624A..64O}. The formation environment of 3I might share similarities with that of C/2016 R2, favouring dinitrogen to ammonia.} 

The N$_2$/CO ratio stored in cometary ices  is important as it can be used to estimate the formation temperature of a comet. Experiments indicate that the trapping efficiency of N$_2$ and CO in amorphous ice is dependent on temperature and different for the two species \citep{2007Icar..190..655B}. The ratio of N$_2$/CO in comets, and how much it departs from the protosolar value, can thus be used to estimate the formation temperature of comets.
\citet{2015Sci...348..232R} measured the N$_2$/CO ratio in comet 67P from in-situ mass spectroscopy data. They found a value of $\sim$0.006 and, {\color{black} assuming a N/C protosolar ratio of 0.29 ± 0.10} \citep{lodders2010solar} {\color{black}entirely converted into $\{$N$_2$/CO\}$_{\odot}$=0.15±0.05, they deduced that the comet must have formed at temperatures lower than 30K.} From measurements later during the Rosetta mission, \citet{2019MNRAS.489..594R} found a N$_2$/CO of 0.03, which indicated an even lower formation temperature closer to 20K. Following a similar reasoning as for 67P, {\color{black}this would indicate a similarly low formation temperature for 3I, which would agree with indications from the hydrogen isotopic ratio in water \citep{2026arXiv260307026S,2026arXiv260306911C}.}

\subsubsection{Main volatiles H$_2$O, CO, CO$_2$}

{\color{black}The neutrals H$_2$O, CO$_2$ and CO can be probed via spectral features in the near infrared or radio, or for H$_2$O estimated via emissions H of OH (UV, radio). Figure 4 of \citet{2026arXiv260708603B} compiles measurements of the production rates of these species in 3I from different instruments. From the sparse measurements across the orbit, they determine trends in the abundance ratios of neutrals with heliocentric distance, which we will use to inform our ion measurements at r$_h$=1.8au.

Measurements of H$_2$O in 3I from different instruments tend to disagree, likely due to an extended source in the form of icy grains. Still they agree on the enrichment of CO compared to H$_2$O in 3I, with H$_2$O/CO systematically smaller than 13. In comparison \citet{2016Icar..278..301D} found an average H$_2$O/CO of 19$\pm$5 among 24 Solar System comets, with only 6 more enriched in CO than 3I. H$_2$O/CO in 3I is also lower than in the comets observed by \citet{HarringtonPinto2022} at similar $r_h$. 
For CO$_2$, the enrichment observed pre-perihelion \citep{2025ApJ...991L..43C,2026ApJ..1000L..52L} lessened as 3I approached the Sun, becoming more typical near perihelion with CO/CO$_2$$\sim$1 and H$_2$O/CO$_2$$\sim$10, then CO-dominated post-perihelion. The H$_2$O/CO$_2$ near perihelion (r$_h$=1.36au) is consistent with the abundances reported by \citet{HarringtonPinto2022} for other comets at similar heliocentric distances. \citet{2012ApJ...752...15O} also reports H$_2$O/CO$_2$ ratios between 3 and 25 for comets these distances.

At r$_h$=1.8au, the trend from \citet{2026arXiv260708603B} indicates H$_2$O/CO$\sim$5. As a first approximation, H$_2$O$^+$/CO$^+$ is expected to be smaller than the ratio of neutrals H$_2$O/CO due the branching ratio of H$_2$O to H$_2$O$^+$ being only 6\% of that of CO to CO$^+$ \citep{1985phae.book..438H,1997A&A...319..617J}. 
However the chemistry of the production and destruction of these ions is complex, and H$_2$O$^+$/CO$^+$ has been observed to vary along the tail \citep[e.g.][]{1997A&A...319..617J}, therefore it is hard to directly link the measured ion ratio and the ratio of neutrals. Using the H$_2$O/CO estimate and the ionized fraction, we expect a produced ion ratio H$_2$O$^+$/CO$^+$$\sim$0.06$\times$5=0.3, which is in good relatively agreement with our measured value H$_2$O$^+$/CO$^+$=0.35$\pm$0.02.
\citet{1993ApJ...407..402L} reports H$_2$O$^+$/CO$^+$ in the plasma tails of 5 comets, all 2 to 10 times larger than our measurement, which is also consistent with the enrichment in CO$^+$ precursors compared to water. 

At r$_h$=1.8au, the trend also estimates H$_2$O/CO$_2$$\sim$10 and CO$_2$/CO$\sim$0.7. Our measurements in 3I (CO$_2$$^+$/CO$^+$=1.4$\pm$0.1, CO$_2$$^+$/H$_2$O$^+$=3.8$\pm$0.4) are close to those made in C/2016 R2 by \citet{2019A&A...624A..64O}: CO$_2$$^+$/CO$^+$=1.1$\pm$0.3 and CO$_2$$^+$/H$_2$O$^+$$>$2.75 at $r_{h}$=$2.7$au. At this larger distance the activity of C/2016 R2 was dominated by CO and CO$_2$. \citet{2019AJ....158..128M} found that it had CO$_2$/CO=0.18$\pm4$, slightly lower than 3I during our observations, and water significantly depleted at H$_2$O/CO$_2$=0.031$\pm$0.004}.

\subsubsection{CH$^+$} \citet{1993ApJ...407..402L} report measurements of the CH$^+$/H$_2$O$^+$ and CH$^+$/CO$^+$ ratios in the tails of comets C/1987 P1 (Bradfield), C/1973 E1 (Kohoutek) and 1P/Halley. 
For CH$^+$/H$_2$O$^+$ they find 0.12$\pm$0.08, 0.03$\pm$0.02 and 0.09$\pm$0.06 in the three comets respectively. This range of values is consistent with our measurement of 0.072$\pm$0.007.
For CH$^+$/CO$^+$ they find 0.086$\pm$0.077, 0.075$\pm$0.063 and 0.15$\pm$0.125 in the three comets respectively. These values are higher than our measurement of 0.025$\pm$0.002. \citet{2026arXiv260307718Z} finds a typical abundance of CH in 3I compared to other minor volatiles. Our measurements of the ion ratios compared to these other comets, assuming a typical production of CH$^+$, agree with the enrichment in CO$_2$  and CO producing CO$^+$.

\subsection{Evolution with distance along the tail}
\label{sec:radial}

While we sampled a large section of the tail for our analysis, it is important to note that the abundance ratios between ions can vary with distance to the comet. Indeed, the different ions have more or less efficient production/loss mechanisms. For example, the photochemical models of \citet{1982A&A...108..221B} usually predict H$_2$O$^+$ column densities decreasing faster with distance than other ions, but they do not consider solar wind interactions from which the tail arise, describing a symmetrical coma instead. They also do not consider hypervolatile rich compositions such as exhibited by 3I. 

Using the data from December 02, where the ion emissions look the most like a well spatially defined tail, we divide the initial aperture into two smaller apertures: the half that is closest to the nucleus ($\sim5\times10^{4}$ to $\sim8\times10^{4}$km) and the half that is farther along the tail ($\sim8\times10^{4}$ to $\sim1.1\times10^{5}$km), although we are seeing the tail from a $\sim30^{\circ}$ angle and these distances do not account for how broad the tail is. In order to assess spatial variations of the abundance ratios along the tail, we calculate the abundance ratios in each small aperture with the methods previously described, and calculate the percentage of increase in the aperture that is farther away from the nucleus compared to the aperture that is closer to the nucleus. 

The results are given in Table \ref{tab:spatial increase}. Over these distances in the plasma tail, given our uncertainties, we do not detect any significant variation of the abundance ratios involving only N$_2$$^+$, CO$^+$, CO$_2$$^+$ or H$_2$O$^+$. This is particularly convenient for N$_2$$^+$/CO$^+$, as it reinforces the accuracy of our measurement.
However, the amount of CH$^+$ seems to be decreasing against every other species, with some of the abundance ratio varying slightly above 1-$\sigma$ level. While these detections are marginal, \citet{1987ApJ...315L.147L} also detected a decrease of CH$^+$/H$_2$O$^+$ in the tail of 1P/Halley between $2\times10^{4}$km and $2\times10^{5}$km when the comet was at $r_{h}$=1.3au, and they could not constrain changes in other ions. \citet{1982A&A...108..221B} do not report on the behaviour of CH$^+$ in their photochemical models. 
Observations and models to compare with are scarce as some of these ions are rarely detected, and their behaviour is dependent on the exhaustive composition of each comet, their activity and environment. The ability to detect and analyse the spatial distribution of these species around comets, as demonstrated in the case of 3I, offers promising perspectives to understand the properties of ion tails. 

\begin{table}
\begin{tabular}{|c|c|c|c|c|c|}
\hline
& A=N$_2$$^+$ & A=CO$^+$ & A=CO$_2$$^+$ & A=H$_2$O$^+$ & A=CH$^+$ \\ \hline
B=N$_2$$^+$  &  0 & - & - & - & -  \\ \hline
B=CO$^+$  &  1.9$\pm$8.4 & 0 & -& - & -  \\ \hline
B=CO$_2$$^+$  &  5$\pm$23 & 3$\pm$21 & 0 & - & - \\ \hline
B=H$_2$O$^+$  &  2.9$\pm$10 & 0.92$\pm$8 & -2$\pm$21 & 0 & -  \\ \hline
B=CH$^+$  &  \textbf{18$\pm$16} & \textbf{16$\pm$14} & 13$\pm$27 & \textbf{15$\pm$15} & 0  \\ \hline
    \end{tabular}
    \caption{Measured increase percentage of the abundance ratios A/B from the aperture close to the nucleus to the one farther away, i.e. \{A/B\}$_{far}$=(1+$p$/100)*\{A/B\}$_{close}$ where $p$ is the tabulated value. Variations larger than 1$\sigma$ are highlighted in bold.}
    \label{tab:spatial increase}
\end{table}

\section{Conclusion}

Using post-perihelion observations of 3I/ATLAS made with WEAVE-LIFU, we determined  abundance ratios between N$_{2}$$^+$, CO$^+$, CO$_{2}$$^+$, H$_{2}$O$^+$ and CH$^+$ ions in the plasma tail of 3I/ATLAS , using an aperture probing distances ranging $\sim$5$\times10^4$km to $\sim$1.1$\times10^5$km along the tail.
Using the N$_{2}^{+}$/CO$^+$ ratio we constrain N$_{2}$/CO$>$0.023$\pm$0.001 in the ices of 3I/ATLAS which, by analogy with Solar System comets and laboratory measurements, seems to indicate cold formation conditions, in agreement with other isotopic evidence.
Ion ratios of other ions are harder to interpret, as they cannot directly be linked to the abundances of neutrals due to the complex coma chemistry, but they appear to be consistent with the enrichment in CO and CO$_{2}$ observed by infrared instruments.
Along the tail, as the distance to the comet increases, the CH$^+$ abundance is found to decrease marginally relatively to all other species, while we do not detect changes in the abundance ratios of other ions.

There are very few comparison points for most of these measurements, whether based on observations of other comets or modelling. These ions are usually hard to detect, due to the limitations of standard instruments, observing geometries which can often be unfavourable for detecting the tail, or typical low abundances of these ions.
Powerful IFUs sensitive to the blue end of the optical spectrum, such as WHT/WEAVE-LIFU, offer new prospects for the remote study of plasma tail composition in addition to being able to probe neutral volatiles and dust. Such instruments will be a precious resource for surveying future comets.

\section*{Acknowledgements}

L.F. is supported by the UKRI Future Leaders Fellowship COMMIT (UKRI2308: COMMIT). 

Funding for the WEAVE facility has been provided by UKRI STFC, the University of Oxford, NOVA, NWO, Instituto de Astrofísica de Canarias (IAC), the Isaac Newton Group partners (STFC, NWO, and Spain, led by the IAC), INAF, CNRS-INSU, the Observatoire de Paris, Région Île-de-France, CONACYT through INAOE, the Ministry of Education, Science and Sports of the Republic of Lithuania, Konkoly Observatory (CSFK), Max-Planck-Institut für Astronomie (MPIA Heidelberg), Lund University, the Leibniz Institute for Astrophysics Potsdam (AIP), the Swedish Research Council, the European Commission, and the University of Pennsylvania.  The WEAVE Survey Consortium consists of the ING, its three partners, represented by UKRI STFC, NWO, and the IAC, NOVA, INAF, GEPI, INAOE, Vilnius University, FTMC – Center for Physical Sciences and Technology (Vilnius), and individual WEAVE Participants. Please see the relevant footnotes for the WEAVE website\footnote{\url{https://weave-project.atlassian.net/wiki/display/WEAVE}} and for the full list of granting agencies and grants supporting WEAVE\footnote{\url{https://weave-project.atlassian.net/wiki/display/WEAVE/WEAVE+Acknowledgements}}.

\section*{Data Availability}

The data underlying this article will be shared on reasonable request to the corresponding author.

\bibliographystyle{mnras}
\bibliography{weave3i} 

\appendix

\section{Abundance ratios at each separate epoch}
\label{sec:appendix}

\begin{table}
\begin{tabular}{|c|c|c|c|c|c|}
\hline
\multicolumn{6}{|c|}{\textbf{Abundance ratios A/B [\%] on 2025 November 30}} \\\hline
    & A=N$_2$$^+$ & A=CO$^+$ & A=CO$_2$$^+$ & A=H$_2$O$^+$ & A=CH$^+$ \\ \hline
B=N$_2$$^+$  &  100 & 4500$\pm$400 & 6300$\pm$800 & 1700$\pm$230 & 110$\pm$10  \\ \hline
B=CO$^+$  &  2.2$\pm$0.2 & 100 & 140$\pm$13 & 38$\pm$4 & 2.4$\pm$0.2  \\ \hline
B=CO$_2$$^+$  &  1.6$\pm$0.2 & 71$\pm$7 & 100 & 27$\pm$4 & 1.7$\pm$0.2  \\ \hline
B=H$_2$O$^+$  &  6$\pm$1 & 270$\pm$30 & 380$\pm$50 & 100 & 6.5$\pm$0.9  \\ \hline
B=CH$^+$  &  92$\pm$10 & 4100$\pm$300 & 5800$\pm$700 & 1500$\pm$200 & 100  \\ \hline
\multicolumn{6}{|c|}{\textbf{Abundance ratios A/B [\%] on 2025 December 02}} \\ \hline
    & A=N$_2$$^+$ & A=CO$^+$ & A=CO$_2$$^+$ & A=H$_2$O$^+$ & A=CH$^+$ \\ \hline
B=N$_2$$^+$  &  100 & 4200$\pm$300 & 5600$\pm$900 & 1500$\pm$100 & 110$\pm$10  \\ \hline
B=CO$^+$  &  2.4$\pm$0.2 & 100 & 130$\pm$20 & 34$\pm$8 & 2.7$\pm$0.3  \\ \hline
B=CO$_2$$^+$  &  1.8$\pm$0.3 & 75$\pm$11 & 100 & 26$\pm$4 & 2$\pm$0.3  \\ \hline
B=H$_2$O$^+$  &  6.9$\pm$0.5 & 290$\pm$20 & 390$\pm$60 & 100 & 7.9$\pm$0.9  \\ \hline
B=CH$^+$  &  88$\pm$9 & 3700$\pm$400 & 5000$\pm$800 & 1300$\pm$200 & 100  \\ \hline

\end{tabular}
\caption{Measured abundance ratios of species A/B given in percents for 2025 November 30 and 2025 December 02.}

\label{tab:resultsappendix}
\end{table}

\bsp	
\label{lastpage}
\end{document}